\documentstyle[12pt]{article}

\begin{document}

\title{\bf {Particle creation by moving spherical shell in the
dynamical Casimir effect }}
\author{
M. R. Setare $^1$ \footnote{E-mail: Mreza@physics.sharif.ac.ir}
 and A. A. Saharian  $^{2}$\footnote{E-mail: saharyan@www.physdep.r.am }  \\
 {$^1$ Department of Physics, Sharif University of
Technology, Tehran, Iran}\\ and \\
{$^{2}$}Department of Physics, Yerevan State University, Yerevan,
Armenia }
\date{\small{\today}}
 \maketitle
\begin{abstract}
The creation of massless scalar particles from the quantum vacuum
by spherical shell with time varying radius is studied. In the
general case of motion the equations are derived for the
instantaneous basis expansion coefficients. The examples are
considered when the mean number of particles can be explicitly
evaluated in the adiabatic approximation.

 \end{abstract}

\newpage

 \section{Introduction}
   The Casimir effect is one of the most interesting manifestations
  of nontrivial properties of the vacuum state in a quantum field
  theory (for reviews see \cite{mueller,Moste,Milt,Birrell}) and
   can be viewed
  as a polarization of
  vacuum by boundary conditions. A new phenomenon, a
  quantum creation of particles (the dynamical Casimir effect)
  occurs when the geometry of the system varies in time. In two
  dimensional spacetime and for conformally invariant fields the
  problem with dynamical boundaries can be mapped to the
  corresponding static problem and hence allows a complete study
  (see \cite{Moste,Birrell} and references therein). In higher
  dimensions the problem is much more complicated and is solved
  for some simple geometries. The vacuum stress induced by uniform
  acceleration of a perfectly reflecting plane is considered in
  \cite{Cand}. The corresponding problem for a sphere
  expanding in the four-dimensional spacetime with constant
  acceleration is investigated by Frolov and Serebriany
  \cite{Frol1,Frol2} in the perfectly reflecting case and by Frolov
  and Singh \cite{Frol3} for semi-transparent boundaries. For more
  general cases of motion by vibrating cavities the problem of particle and
  energy creation is considered on the base of various
  perturbation methods \cite{Caluc, Hacyan,Widom,dod,Lamb,Ji,Schut,dod1}(for more
  complete list of references see \cite{dod1}).
   It have been
  shown that a gradual accumulation of small changes in the quantum
  state of the field could result in a significant observable
  effect. A new application of the dynamical Casimir effect has
  recently appeared in connection with the suggestion by Schwinger
  \cite{Schwing} that the photon production associated with changes in
  the quantum electrodynamic vacuum state arising from a
  collapsing dielectric bubble could be relevant for
  sonoluminescence (the phenomenon of light
  emission by a sound-driven gas bubble in a fluid \cite{Bar}).
  For the further developments and discussions this quantum-vacuum
  approach see \cite{Milton1,Eber,lib,MilNg,Liber2} and references
  therein. In the present paper we consider particle creation from
  the quantum scalar vacuum by expanding or contracting spherical
  shell with Dirichlet boundary conditions. In next section we
  derive the equations for the instantaneous basis expansion
  coefficients and the formula for the Bogoliubov coefficients.
  The examples when this coefficients can be explicitly found at
  adiabatic approximation are discussed in section 3.

\section{General consideration}

Consider a scalar field $\varphi $ satisfying Dirichlet boundary
condition on the surface of a sphere with time-dependent radius
$a=a(t)$:
\begin{equation} \label{fieldeq}
(\frac{\partial^{2}}{\partial t^{2}}-\Delta)\varphi (x,t)=0,\qquad
\varphi |_{r=a(t)}=0.
\end{equation}
The corresponding eigenfunctions can be expanded in a series with
respect to the instantaneous basis
\begin{equation}\label{eigenfunction}
\varphi _{\alpha}(x)=\sum_{\beta }q_{\alpha \beta }(t)\phi
_{\beta }(x,a(t)),
\end{equation}
where the collective index $\alpha$ denotes the set of quantum
numbers specifying the solution. Here
$\phi_{\beta}(x,a)\exp(-\imath \omega t)$ are the corresponding
eigenfunctions for a static sphere with radius $a$:
\begin{eqnarray} \label{eigst}
\phi_{\beta }(x,a)&=&\frac{\sqrt{2}j_{l}(j_{l,n}r/a)} {a^{3/2}j'_{l}(j_{l,n})}Y_{lm}(\theta ,\varphi ), \\
\beta &=&(l,m,n), \quad l=0,1,2,..., \quad -l\leq m\leq l, \quad
n=1,2,..., \nonumber
\end{eqnarray}
with $j_{l,n}$ being the n-th zero for the spherical Bessel
function $j_{l}(z)$, $j_{l}(j_{l,n})=0$, $Y_{lm}(\theta ,\varphi
)$ is the spherical harmonic. Note that the sets of quantum
numbers are the same for static and dynamic cases. Function
(\ref{eigenfunction}) satisfies boundary condition
(\ref{fieldeq}). Putting expression (\ref{eigenfunction}) into
the field equation (\ref{fieldeq}) we obtain
\begin{equation}\label{eq1}
\sum_{\beta }\left\{ \phi _{\beta }[\ddot{q}_{\alpha \beta
}(t)+\omega _{\beta }^{2}(t) q_{\alpha\beta}(t)]+2\dot{q}_{\alpha
\beta }\dot{\phi }_{\beta }+ q_{\alpha \beta}\ddot{\phi }_{\beta
}\right\} =0 ,\quad \omega_{\beta }(t)=j_{l,n}/a(t),
\end{equation}
where dot stands for the time-derivative. Let us multiply this
equation by $\phi_{\beta'}^{\ast}$ and integrate over the region
inside a sphere at a given moment $t$. Using the orthonormality
relation
\begin{equation}\label{ortrel}
\int \phi_{\beta}\phi_{\beta'}^{\ast}d^{3}x=\delta_{\beta\beta'},
\end{equation}
this yields
\begin{equation}\label{qeq2}
\ddot{q}_{\alpha\beta}(t)+\omega_{\beta }^{2}(t)
q_{\alpha\beta}(t)=\sum_{\beta'}(2\dot{q}_{\alpha \beta '}g_{\beta
'\beta}+ q_{\alpha\beta '}g_{\beta '\beta}^{(1)}),
\end{equation}
where we have introduced notations
\begin{equation}\label{gg1}
g_{\beta'\beta}=-\int \dot{\phi
}_{\beta'}\phi_{\beta}^{\ast}d^{3}x, \hspace{1cm}
g_{\beta'\beta}^{(1)}=-\int \ddot{\phi }
_{\beta'}\phi_{\beta}^{\ast}d^{3}x,
\end{equation}
with integrations over the region inside the sphere. The relation
between coefficients (\ref{gg1}) can be found by making use the
completeness condition for eigenfunction (\ref{eigst}):
\begin{equation}\label{compcond}
\sum_{\gamma}\phi_{\gamma}(\vec{r})\phi_{\gamma}^{*}(\vec{r}')=\delta
(\vec{r}-\vec{r}').
\end{equation}
This yields
\begin{equation}\label{relg1g}
g_{\beta'\beta}^{(1)}=\frac{\partial g_{\beta'\beta}}{\partial
t}+\sum_{\gamma}g_{\beta'\gamma}g_{\beta\gamma}^{*}.
\end{equation}
By taking into account this relation we note that the general
structure of equations (\ref{qeq2}) for the instantaneous basis
expansion coefficients is similar to that for the plane case.
 Using the standard orthonormality relations for the spherical
harmonic it can be easily seen that
\begin{equation}\label{gg2}
g_{\beta'\beta}=g_{n'n}^{l}\delta_{ll'}\delta_{mm'},\hspace{2cm}
g_{\beta'\beta}^{(1)}=g_{n'n}^{(1)l}\delta_{ll'}\delta_{mm'}.
\end{equation}
To evaluate the integrals in (\ref{gg1}) we use the formula for
the integrals involving the Bessel spherical function
\begin{equation} \label{Besint}
\int_{0}^{1}z^{2}j_{l}(az)j_{l}(bz)dz=\frac{aj'_{l}(a)j_{l}(b)-
bj_{l}(a)j'_{l}(b)}
{b^{2}-a^{2}}.
\end{equation}
From this formula the following expression for the coefficients
can be obtained
\begin{equation}\label{glnn}
g_{n'n}^{l}=h a_{n'n}^{l}, \quad g_{n'n}^{(1)l}= \dot{h}
a_{n'n}^{l}+h^{2}\sum_{p}a_{n'p}^{l}a_{np}^{l},\hspace{1cm}
h=\frac{\dot{a}}{a},
\end{equation}
where
\begin{eqnarray}\label{geq}
a_{nn}^{l}& =0, \\
a_{n'n}^{l}& =& \frac{2j_{l,n'}j_{l,n}}{j_{l,n'}^{2}-
j_{l,n}^{2}},\hspace{1cm}n\neq n'. \nonumber
\end{eqnarray}
As we see the coefficients $g_{\beta'\beta'}$ are antisymmetric.
This is a direct consequence of orthonormality relation
(\ref{ortrel}).
 From expressions (\ref{gg1}) it follows that the coefficients
$q_{\alpha\beta}$ in (\ref{eigenfunction}) can be chosen diagonal
with respect to the quantum numbers $l$ and $m$, and are
independent on $m$:
\begin{equation} \label{qdiag}
q_{\beta\beta'}=\delta_{ll'}\delta_{mm'}q_{nn'}^{l}.
\end{equation}
Now the expansion for the eigenfunctions takes the form
\begin{equation} \label{neweig}
\varphi_{lmn}=\sum_{n'=1}^{\infty}q_{nn'}^{l}\phi_{lmn'}(x,a(t)),
\end{equation}
with the following infinite set of coupled differential equations
for the coefficients:
\begin{equation} \label{eqcoef}
\ddot{q}_{nn'}^{l}+\omega_{ln'}^{2}(t) q_{nn'}^{l}=2h
\sum_{p=1}^{\infty}\dot{q}_{np}^{l} a_{pn'}^{l}+ \dot{h}
\sum_{p=1}^{\infty}q_{np}^{l} a_{pn'}^{l}+ h^{2}
\sum_{p,s=1}^{\infty} q_{np}^{l}a_{ps}^{l}a_{n's}^{l}.
\end{equation}
If the sphere is asymptotically static at past and future then
the in- and out- vacuum states can be defined by using the
solutions for coefficients corresponding to the in- and out-modes
$\varphi _{\alpha}^{{\rm (in)}}(t), \, \varphi _{\alpha}^{{\rm
(out)}}(t)$ with asymptotics
\begin{eqnarray} \label{qinout}
q_{\alpha \beta }^{{\rm (in)}}(t)&\rightarrow &e^{-\imath
\omega_{\alpha }^{{\rm in}}t}\delta _{\alpha \beta },\quad
t\rightarrow -\infty  \\
q_{\alpha \beta }^{{\rm (out)}}(t)&\rightarrow &e^{-\imath \omega
^{{\rm out}}_{\alpha }t}\delta _{\alpha \beta }, \quad
t\rightarrow \infty ,
\end{eqnarray}
where we use the notations
\begin{equation} \label{omegalf}
\omega_{\alpha }^{{\rm in}}=\frac{j_{l,n}}{a_{-}},  \qquad
\omega_{\alpha }^{{\rm out}}=\frac{j_{l,n}}{a_{+}}, \qquad
a_{\pm}=\lim_{t\rightarrow \pm \infty}a(t)
\end{equation}
for the corresponding eigenfrequencies. The field operator in the
Heisenberg representation may be expanded in terms of the
corresponding eigenfunctions
\begin{equation} \label{phiinout}
\varphi _{\alpha}^{(s)}=\sum_{\beta }q_{\alpha \beta
}^{(s)}(t)\phi_{\beta }(x,a(t)),\hspace{1cm} s={\rm in,out}
\end{equation}
as
\begin{equation} \label{phiexp2}
\varphi
(x,t)=\sum_{\alpha}[a_{\alpha}^{(s)}\varphi_{\alpha}^{(s)}+
a_{\alpha}^{(s)+}\varphi_{\alpha}^{(s)\ast}].
\end{equation}
The in and out vacuum states $|{\rm in}>,|{\rm out }>$ are defined
in accordance with
\begin{equation} \label{vac2}
a_{\alpha}^{(s)}|s>=0.
\end{equation}
The corresponding eigenfunctions are related by the Bogoliubov
transformation
\begin{equation}\label{Bogtrans}
\varphi_{\alpha}^{{\rm
(in)}}=\sum_{\beta}(\alpha_{\alpha\beta}\varphi_{\beta}^{{\rm
(out)}} +\beta_{\alpha\beta}\varphi_{\beta}^{{\rm (out)}\ast}),
\end{equation}
with Bogoliubov coefficients $\alpha_{\alpha\beta}$ and
$\beta_{\alpha\beta}$. Substituting instantaneous basis expansion,
multiplying by $\varphi_{\gamma}^{\ast}$ and integrating over the
region inside the sphere we obtain the corresponding relation
between expansion coefficients:
\begin{equation}
q_{\alpha\gamma}^{{\rm
(in)}}(t)=\sum_{\beta}(\alpha_{\alpha\beta}q_{\beta\gamma} ^{{\rm
(out)}}+ \beta_{\alpha\beta}q_{\beta\gamma}^{{\rm (out)}\ast}).
\label{Bogq}
\end{equation}
Here we used the formula
\begin{equation}\label{int1}
\int
\phi_{\beta}^{\ast}\phi_{\beta'}^{\ast}d^{3}x=\delta_{nn'}\delta_{ll'}
\delta_{m,-m'},
\end{equation}
and the fact that expansion coefficients are independent on $m$.
Now by taking into account relation (\ref{qdiag}) the Bogoliubov
transformation can be written in the form
\begin{equation}
q_{nn'}^{{\rm (in)} l }(t)=\sum_{p}(\alpha_{np}^{l}q_{pn'} ^{{\rm
(out)} l}+ \beta_{np}^{l}q_{pn'}^{{\rm (out)}l\ast }).
\label{qinBog}
\end{equation}
In particular, taking into account (\ref{qinout}), in the limit
$t\rightarrow\infty$ from (\ref{Bogq}) we receive
\begin{equation}
q_{\alpha\gamma}^{{\rm
in}}(t)=\alpha_{\alpha\beta}e^{-i\omega_{\beta}^{{\rm out}}t}+
\beta_{\alpha\beta}e^{i\omega_{\beta}^{{\rm out}}t},\qquad
t\rightarrow +\infty . \label{Bogtinf}
\end{equation}
 The mean number of out particles with quantum number $\alpha$
in the in-vacuum state is determined by the Bogoliubov coefficient
$\beta_{\sigma\alpha}$ as
\begin{equation} \label{Npart}
<{\rm in}|N_{\alpha}|{\rm
in}>=\sum_{\sigma}|\beta_{\sigma\alpha}|^{2},\qquad
\beta_{\sigma\alpha}=-(\varphi _{\sigma}^{{\rm (in)}},\varphi
_{\alpha}^{{\rm (out)}*}),
\end{equation}
with the standard Klein-Gordon scalar product (see, for instance,
\cite{Birrell}). From (\ref{Npart}) it follows that
$\beta_{\alpha'\alpha}=\beta_{n'n}^{l}\delta_{ll'}\delta_{m,-m'}$
and hence the total number of created scalar particles is given by
\begin{equation} \label{Ntot}
<{\rm in}|N|{\rm in}>=\sum_{lmn}<{\rm in}|N_{lmn}|{\rm
in}>=\sum_{l=0}^{\infty}(2l+1)\sum_{n,n'=1}^{\infty}|\beta_{n'n}^{l}|^{2}
\end{equation}
The coefficients $\alpha_{np}^{l}$ and $\beta_{np}^{l}$ in
(\ref{Bogq}) are determined from the solutions of infinite set of
coupled differential equations (\ref{eqcoef}) with time dependent
coefficients. The problem can be simplified in limiting cases
when various approximations can be used (see, for example,
\cite{Caluc}-\cite{dod1} for the case of plane boundaries). In
next section we consider the adiabatic approximation.

\section{Adiabatic approximation }

From the form of equations (\ref{eqcoef}) it follows that there
are two types of effects which lead to the particle creation (see
also \cite{Schut}). The first one, called squeezing of the
vacuum, is due to the nonstationary eigenfrequencies
$\omega_{ln}(t)$ as a result of a dynamical change of the radius
of the sphere and is described by the second term on the left of
(\ref{eqcoef}). The second one, referred as acceleration effect,
is due to the motion of the boundary and comes from the terms on
the right of (\ref{eqcoef}). Due to the antisymmetry of the
coefficients $a_{nn'}^{l}$ the squeezing- and acceleration-
effects give additive contributions to the number of created
particles per mode to first non-vanishing order of perturbation
theory. In this section we will consider the squeezing
contribution to the number of particles. Note that when the
sphere is moving on a time scale slow on the scale of the created
quanta frequencies (adiabatic approximation)(see \cite{Widom} and
references therein) the terms on the right of (\ref{eqcoef})
containing the derivatives of the sphere radius are small and the
squeezing effect is dominant. In the adiabatic approximation the
matrix $q_{\alpha\beta}$ can be chosen in diagonal form
$q_{\alpha\beta}=q_{\alpha}\delta_{\alpha\beta}$ and
$q_{nn'}^{l}=q_{ln}\delta_{nn'}$ and from (\ref{eqcoef}) we
receive
\begin{equation}\label{eqad}
\ddot{q}_{ln}(t)+\omega_{ln}^{2}(t)q_{ln}(t)=0.
\end{equation}
As it follows from (\ref{eqcoef}) the necessary conditions for the
adiabatic approximation are
\begin{equation}
\dot{a}^{2}, \quad a\ddot{a}\ll j_{l,n}^{2} \label{adcond}.
\end{equation}
As an example let us consider an exactly solvable case when
\begin{equation}
a(t)=\frac{1}{\sqrt{A+ B \tanh(t/t_{0})}},\hspace{1cm}A>|B|,
\label{exth}
\end{equation}
where $A,B$ and $t_{0}$ are constants. This motion corresponds to
the sphere contraction for $ B>0$ and expansion for $B<0$. The
corresponding frequencies are
\begin{eqnarray}\label{freqth}
\omega_{ln}(t)&=& j_{l,n}\sqrt{A+B\tanh(t/t_{0})}, \\
\omega^{{\rm in}}_{ln}&=& \frac{j_{l,n}}{a(-\infty)}=
j_{l,n}\sqrt{A-B},\qquad \omega^{{\rm out}}_{ln}=
\frac{j_{l,n}}{a(\infty)}= j_{l,n}\sqrt{A+B}.
\end{eqnarray}
Now we need to solve the equation (\ref{eqad}) with
$\omega_{l,n}(t)$ given by (\ref{freqth}). The corresponding
solutions are given by hypergeometric function. The normalized
in- and out- modes are given by formula \cite{Birrell}
 \begin{eqnarray} \label{hypgeom}
q^{s}_{ln}(t)&=&(2\omega^{s}_{ln})^{-1/2}\exp[-\imath\omega_{ln}^{+}t-
\imath\omega_{ln}^{-}t_{0} \ln[2\cosh(t/t_{0})]] \times \nonumber \\
&\times &  \, {}_{2}F_{1}(1+\imath\omega_{ln}^{-}t_{0},
\imath\omega_{ln}^{-}t_{0};1\mp
\imath\omega_{ln}^{s}t_{0};\frac{1}{2}(1\pm\tanh(t/t_{0}))),\quad
s={\rm in},\, {\rm out} ,
\end{eqnarray}
where uper/lower sign corresponds to the in/out- modes, and
\begin{equation} \label{omegpm}
\omega ^{\pm}_{ln}=\frac{1}{2}(\omega_{ln}^{{\rm out}}\pm
\omega_{ln}^{{\rm in}}).
\end{equation}
Using the diagonality for the coefficients $q_{nn'}^{l}$ over
$nn'$ we conclude that the same is the case for Bogoliubov
coefficients in (\ref{qinBog}):
 \begin{equation}\label{alfdiag}
 \alpha_{nn'}^{l}=\alpha_{ln}\delta_{nn'},\hspace{1cm}\beta_{nn'}^{l}=
 \beta_{ln}\delta_{nn'}.
 \end{equation}
 Now the Bogoliubov transformation (\ref{qinBog}) can be written
 as
 \begin{equation}\label{qinBogdiag}
q^{{\rm (in)}}_{ln}=\alpha_{ln} q^{{\rm (out)}}_{ln}+\beta_{ln}
q^{{\rm (out)} *}_{ln}.
\end{equation}
Using the linear relation between hypergeometic functions,
similar to \cite{Birrell} for the coefficients in this formula
one finds
\begin{equation} \label{Bogalf3}
\alpha_{ln}=\left( \frac{\omega_{ln}^{{\rm
out}}}{\omega_{ln}^{{\rm in}}}\right) ^{1/2}
\frac{\Gamma(1-\imath\omega_{ln}^{{\rm in}}t_{0})
\Gamma(-\imath\omega_{ln}^{{\rm out}}t_{0})}
{\Gamma(-\imath\omega_{ln}^{+}t_{0})
\Gamma(1-\imath\omega_{ln}^{+}t_{0}))},
\end{equation}

\begin{equation} \label{Bogbet3}
\beta_{ln}=\left( \frac{\omega_{ln}^{{\rm
out}}}{\omega_{ln}^{{\rm in}}}\right) ^{1/2}
\frac{\Gamma(1-\imath\omega_{ln}^{{\rm
in}}t_{0})\Gamma(\imath\omega_{ln}^{{\rm out}}t_{0})}
{\Gamma(\imath\omega_{ln}^{-}t_{0})\Gamma(1+
\imath\omega_{ln}^{-}t_{0})}.
\end{equation}
The mean number of particles produced through the modulation of
the single scalar mode is
\begin{equation} \label{Nex1}
<{\rm in}|N_{ln}|{\rm
in}>=|\beta_{ln}|^{2}=\frac{\sinh^{2}(\pi\omega_{ln}^{-}t_{0})}
{\sinh(\pi\omega_{ln}^{{\rm in}}t_{0})\sinh(\pi\omega_{ln}^{{\rm
out}}t_{0})}.
\end{equation}
The total number of particles produced is obtained by taking the
sum over all the oscillation modes inside the sphere:
\begin{equation} \label{Ntotex1}
<{\rm in}|N|{\rm in}>=
\sum_{l=0}^{\infty}(2l+1)\sum_{n=1}^{\infty} \frac{\sinh^{2}[\pi
j_{l,n}t_{0}(\sqrt{A+B}-\sqrt{A-B})/2]}{\sinh(\pi
j_{l,n}t_{0}\sqrt{A-B})\sinh(\pi j_{l,n}t_{0}\sqrt{A+B})} .
\end{equation}
Therefore the energy related to the particles production inside
the sphere is given by
\begin{equation} \label{enex1}
E=\sum_{l=0}^{\infty}\sum_{n=1}^{\infty} N_{ln}\omega_{l,n}^{{\rm
out}}= \sum_{l=0}^{\infty}\sum_{n=1}^{\infty}(2l+1)
\frac{\sinh^{2}[\pi
j_{l,n}t_{0}(\sqrt{A+B}-\sqrt{A-B})/2]}{\sinh(\pi
j_{l,n}t_{0}\sqrt{A-B})\sinh(\pi j_{l,n}t_{0}\sqrt{A+B})}
j_{l,n}\sqrt{A+B}.
\end{equation}
It can be seen that the similar results may be obtained for
another type of boundary conditions, for instance, for the
Neumann or more general Robin conditions taking instead of
$j_{l,n}$ the corresponding eigenvalues.

 As an another example
let us consider the case in which the sphere radius varies as (for
the corresponding problem in the case of parallel plates see
\cite{Widom})
\begin{equation}\label{pulse}
a(t)=a_{0}\left[ 1+\frac{b^{2}}{\cosh^{2}(t/t_{0})}\right] ^{-1/2}
\end{equation}
with $a_{0},b,t_{0}$ being constants. This function describes a
pulse with characteristic  duration $t_{0}$ and modulation
strength $b$. As in the previous example the corresponding in-
and out-solutions to equation (\ref{eqad}) can be expressed in
terms of hypergeometric function (see, for instance,
\cite{Landau}). Expanding in-modes in terms of the out functions
for the mean number of scalar particles with a given mode one
receives
\begin{equation}\label{npulse}
<{\rm in}|N_{ln}|{\rm in}>=\frac{\cos^{2}[(\pi /2)
\sqrt{1+(2j_{l,n}bt_{0}/a_{0})^{2}}]} {\sinh^{2}(\pi
j_{l,n}t_{0}/a_{0})}.
\end{equation}
An alternative way to obtain this result is  o use the
corresponding quantum mechanical reflection probability given in
\cite{Landau}. Now we can find the total energy released by
adding the energy of each quantum:
\begin{equation}\label{epulse}
E=\frac{1}{a_{0}}\sum_{l=0}^{\infty}(2l+1)\sum_{n=1}^{\infty}j_{l,n}
\frac{\cos^{2}[(\pi /2)\sqrt{1+(2j_{l,n}bt_{0}/a_{0})^{2}}]}
{\sinh^{2}(\pi j_{l,n}t_{0}/a_{0})}.
\end{equation}
Recall that we have obtained formulae (\ref{Nex1}) and
(\ref{npulse}) neglecting the terms on the right of equations
(\ref{eqcoef}), and hence these formulae give the number of
particles produced due to the squeezing effect. In general, the
acceleration effect will give additional contribution to the
particle creation. It is important to stress again that these two
types of effects give additive contributions to first
non-vanishing order of perturbation theory. As a result the
formulae (\ref{Nex1}) and (\ref{npulse}) as giving a part of the
quanta number due to the squeezing effect are valid for more
general situations, when additional contributions to the total
number of particles have to be taken into account due to the
acceleration effect.

 By using the results given above for a scalar
field we can obtain the number of produced particles in the
physically more realistic electromagnetic case. For this note that
electromagnetic field in a spherical cavity is a superposition of
two types of modes: transverse electric (TE) and transverse
magnetic (TM). The TE-modes correspond to the scalar modes with
Dirichlet boundary condition (excluding the $l=0$ mode) and
TM-modes correspond to the scalar modes with mixed type (Robin)
boundary condition
\begin{equation}\label{TMmodes}
j_{l}(\omega a)+\omega a j_{l}'(\omega a)=0,
\end{equation}
on the sphere surface $r=a$. Now the number of photons produced
can be presented in the form
\begin{equation}\label{photon}
<{\rm in}|N_{ln}|{\rm in}>=<{\rm in}|N_{ln}^{{\rm (TE)}}|{\rm
in}>+<{\rm in}|N_{ln}^{{\rm (TM)}}|{\rm in}>, \quad l=1,2,...
\end{equation}
where the first summand on the right is given by formulae
(\ref{Nex1}) and (\ref{npulse}), and the corresponding expressions
for the second summand are obtained from these formulae by
replacement $j_{l,n}\rightarrow \gamma_{l,n}$, where $\omega
a=\gamma_{l,n}$ are solutions to the equation (\ref{TMmodes}).

\section{Conclusion}

In this paper we have considered the particle creation from the
scalar vacuum by moving spherical boundary. For the general case
the set of equations is derived for the instantaneous basis
expansion coefficients and corresponding Bogoliubuv coefficients
are considered. To solve these equations and to obtain the
particle number, various approximations can be used. Here we use
the adiabatic approximation, assuming small velocities for the
sphere motion or large frequencies for the radiation quanta.
Specific examples are considered when the number of particles
produced can be explicitly found. The first one corresponds to
the expansion or contraction of the sphere between two finite
values of the radius, and second one corresponds to the pulse
described by (\ref{pulse}). The results for the electromagnetic
field can be obtained by summing the contributions from TE and TM
modes, corresponding to the scalar modes with Dirichlet and
special Robin type boundary conditions. In general, when the
acceleration effect in the particle creation cannot be neglected
the formulae presented in previous section give only the
squeezing parts of the total number of particles. To first
non-vanishing order of perturbation theory the contributions from
these two types of effects are additive.

\section*{Acknowledgments}

This work has been completed during the stay of one of the
authors (A. A. S.) at Sharif University of Technology. It is a
pleasant duty for him to thank the Department of Physics and
Prof. Reza Mansouri for kind hospitality.

\end{document}